\documentclass[10pt,twocolumn,letterpaper]{article}
\usepackage[pagenumbers]{cvpr}

\usepackage[utf8]{inputenc}
\usepackage[T1]{fontenc}

\usepackage{amsmath}

\usepackage{graphicx}
\usepackage{hyperref}

\usepackage{bm}
\usepackage[separate-uncertainty=true]{siunitx} 

\date{}

\title{Latent Disentanglement in Mesh Variational Autoencoders Improves the Diagnosis of Craniofacial Syndromes and Aids Surgical Planning}

\author{Simone Foti $^{1, 2, *}$, Alexander J. Rickart $^3$, Bongjin Koo $^{1, 2, 4}$, Eimear O' Sullivan $^{3, 5}$, \\ Lara S. van de Lande $^6$, Athanasios Papaioannou $^{3, 5}$, Roman Khonsari $^7$, \\ Danail Stoyanov $^{1, 2}$, N. u. Owase Jeelani $^3$, Silvia Schievano $^3$, \\ David J. Dunaway $^3$, Matthew J. Clarkson $^{1, 2}$ \\[2mm]
\footnotesize $^1$ Wellcome/EPSRC Centre for Interventional and Surgical Sciences, University College London, London, UK \\
\footnotesize $^2$ Centre For Medical Image Computing, University College London, London, UK \\
\footnotesize $^3$ UCL Great Ormond Street Institute of Child Health and Craniofacial Unit, Great Ormond Street Hospital for Children, London, UK \\
\footnotesize $^4$ University of California, Santa Barbara, Department of Electrical \& Computer Engineering, Santa Barbara, USA \\
\footnotesize $^5$ Imperial College London, Department of Computing, London, UK \\
\footnotesize $^6$ Department of Oral and Maxillofacial Surgery, Erasmus Medical Center, Rotterdam, The Netherlands \\
\footnotesize $^7$ Department of Maxillofacial Surgery and Plastic Surgery, Necker - Enfants Malades University Hospital, Paris, France. \\ [1mm]
\small $^*$ s.foti@cs.ucl.ac.uk
}

\begin{document}
    
    \flushbottom
    \maketitle
    
    \begin{abstract}
        \noindent The use of deep learning to undertake shape analysis of the complexities of the human head holds great promise. However, there have traditionally been a number of barriers to accurate modelling, especially when operating on both a global and local level. In this work, we will discuss the application of the Swap Disentangled Variational Autoencoder (SD-VAE) with relevance to Crouzon, Apert and Muenke syndromes. 
        Although syndrome classification is performed on the entire mesh, it is also possible, for the first time, to analyse the influence of each region of the head on the syndromic phenotype. By manipulating specific parameters of the generative model, and producing procedure-specific new shapes, it is also possible to simulate the outcome of a range of craniofacial surgical procedures. This opens new avenues to advance diagnosis, aid surgical planning and allows for the objective evaluation of surgical outcomes.
    \end{abstract}
    
    \section{Introduction}
        \label{sec:intro}
        \noindent Crouzon, Apert and Muenke are craniofacial syndromes characterised by complex craniosynostosis and craniofacial deformity. Although both phenotypically
        and genetically distinct, they are all the result of gain-of-function mutations affecting fibroblast growth factor receptors~\cite{wilkie2017clinical}. Premature fusion of cranial sutures alongside asymmetry and growth restriction at the skull base lead to the development of an abnormal head shape and a risk of developing raised intra-cranial pressure~\cite{calandrelli2014radiological, ruggiero2019syndromic}. Muenke syndrome is the most common genetic abnormality associated with craniosynostosis ($1$ in $10,000$ - $1$ in $30,000$ live births), in contrast to Crouzon ($1$ in $25,000$) and Apert ($1$ in $100,000$)~\cite{johnson2011craniosynostosis}. Crouzon and Apert syndrome are also frequently complicated by airway impairment and shallow orbits which confer little ocular protection. Major craniofacial surgery is often required to overcome these problems, requiring extensive planning and multi-disciplinary expertise~\cite{ruggiero2019syndromic}.

        The initial studies aimed at analysing the shape properties of craniofacial syndromes and their corrective procedures were based on morphometric analysis, where distances, angles, and ratios were the main criteria to compare different shapes \cite{goldberg1981some, cohen1993growth, farkas1994anthropometry}.
        Now, advances in artificial intelligence, machine learning, and deep learning have opened new avenues for shape analysis and for their integration into clinical practice~\cite{o2021craniofacial, meulstee2015new, knoops2019machine, crombag2014assessing, lin2021construction, elkhill2023geometric}. Perhaps the most valuable application of deep learning in plastic and reconstructive surgery lies in virtual surgical planning~\cite{jarvis2020artificial}. 
        Since the first adoptions of Principal Component Analysis (PCA) to analyse head shapes of patients with craniofacial syndromes~\cite{crombag2014assessing,heuze2014quantification}, PCA-based statistical shape models became increasingly common in craniofacial studies \cite{li2015statistical, meulstee2015new, maas2018using, knoops2019machine, o20213d, ortun2020towards}. The low dimensional vectors known as principal mode of variations of the PCA models, were either directly observed and interpolated \cite{heuze2014quantification, crombag2014assessing, meulstee2015new, maas2018using}, or analysed with other methods like SVMs \cite{knoops2019machine, o20213d}, manifold visualisation techniques \cite{knoops2019machine, o20213d}, or linear regression methods \cite{heuze2014quantification, li2015statistical, knoops2019machine, o20213d}.
        
        Particularly relevant are also the multiple techniques aimed at identifying genetic disorders from the shape of the head~\cite{gurovich2019identifying, hallgrimsson2020automated, bannister2022deep, bannister2022detecting, mahdi20223d, mahdi2022multi}. The methods based on mesh autoencoders repeatedly proved their superior performance when compared to PCA baselines~\cite{mahdi20223d,mahdi2022multi}, thus justifying the adoption of mesh autoencoder architectures in future studies.
        Interestingly, some of these methods also offered some level of interpretability by providing some intuition on the facial features used to justify a particular inference~\cite{bannister2022deep}, by displaying colourmaps proportional to the gradients backpropagated on the input shapes~\cite{bannister2022detecting}, or by combining separate models operating on a limited number of facial features~\cite{mahdi2022multi}. 
        More similar to ours, is the work of \cite{o2021craniofacial}, which used mesh autoencoders and support vector machines (SVMs) to diagnose the same craniofacial syndromes analysed also in our work. Since their framework achieved near to perfect classification accuracy, the purpose of our work is to increase the diagnostic performances in terms of interpretability and granularity while introducing surgical simulation capabilities, rather than further improving the classification accuracy. Note that the main downsides to the approach introduced in \cite{o2021craniofacial} are the inability to control the generation of new shapes and to disentangle local shape attributes. In fact, this method can be used only to analyse global shape properties, limiting its adoption in surgical planning and preventing a more granular interpretation of the results with respect to local shape properties.  
            
        Craniofacial surgery has a range of anatomical and technical limitations, and the face is often considered as a group of sub-units by which we can approach the correction of deformity. As such, when considering the translation of machine learning models into the surgical environment, an essential component is the ability to analyse and generate influential regional anatomy whilst still retaining the context of the face as a whole. Traditional generative and shape analysis models, such as statistical shape~\cite{maas2018using, knoops2019machine} and 3D morphable models~\cite{ploumpis2019combining, li2017learning, blanz1999morphable}, autoencoders~\cite{gong2019spiralnet++, cosmo2020limp}, variational autoencoders~\cite{ranjan2018generating, aumentado2019geometric}, and generative adversarial networks~\cite{li2020learning, gecer2020synthesizing, cheng2019meshgan, abrevaya2019decoupled} are all unable to operate on regional anatomy. 
        In fact, it is worth highlighting that the core idea of all the existing models is to learn a mapping between 3D shapes and low-dimensional vectors. Once the mapping is established, shapes can be represented by small vectors and, vice versa, vectors can be transformed into 3D shapes. Learning a mapping between shapes and a low-dimensional space is undeniably convenient, but the learned space is usually difficult to interpret. 
        Recent publication of our underpinning methodology~\cite{foti20223d}, introduced a new approach to train a Variational Autoencoder (VAE) and obtain a more interpretable latent representation that defines direct correspondences between subsets of variables and local shape attributes. The model was originally used to control the generation of 3D faces and bodies for applications in augmented and virtual reality as well as for the movie and video-game production. In this work, we will demonstrate the application of the Swap Disentangled Variational Autoencoder (SD-VAE)~\cite{foti20223d} in diagnosis and objective identification of the key morphological features that contribute to each syndromic phenotype as well as the scope for aiding in planning and the assessment of outcomes in craniofacial surgery. Compared to other models offering some level of interpretability~\cite{bannister2022deep, bannister2022detecting, mahdi2022multi}, the proposed method has the advantage of simultaneously considering the global shape as well as the local anatomical sub-units. Furthermore, it increases the number of regions while still relying on a single network that enables attribute-specific interpolations useful for surgical simulation.
        In addition, to overcome the shortage of data for syndromic patients as well as the significant class imbalance between the different syndromes, we introduce a novel data augmentation technique operating in the spectral domain.        
            
        To summarise, this work provides \emph{(i)} new insights to improve the diagnosis of craniofacial syndromes, \emph{(ii)} more interpretable computational tools that are capable of automatically diagnosing Crouzon, Apert, and Muenke patients, \emph{(iii)} a method to statistically simulate different surgical procedures, and \emph{(iv)} a novel data augmentation technique that mitigates the issues arising from the paucity of 3D meshes representing patients with rare conditions.

    \section{Methods} 
        \label{sec:methods}
        
        \subsection{Data sources and processing.}
            \label{sec:data}
            \noindent We rely on the dataset introduced in \cite{o2021craniofacial}, which includes head meshes of syndromic craniosynostosis patients as well as healthy subjects. The collection of patient data was approved by the UK Research Ethics Committee (UK REC 15/LO/0386) and Great Ormond Street Hospital for Children R\&D office (R\&D 14DS25) and informed consent was obtained from either the patients or their legal guardians~\cite{o2021craniofacial}.
            
            The meshes of syndromic patients diagnosed with Apert, Crouzon, or Muenke syndrome were all segmented from Computed Tomography (CT). Meshes of healthy subjects were obtained from either CT scans or 3D photographs acquired for the creation of the Liverpool-York head model (LYHM)~\cite{dai2020statistical}. 
            The final dataset collected in \cite{o2021craniofacial} includes $39$ Apert patients, $53$ Crouzon patients, $11$ Muenke patients, and $250$ healthy subjects (with $139$ coming from the LYHM data). All subjects are aged between $1$-day and $20$-years-old with a male to female ratio of $55:45$.
            The meshes were put in dense point-correspondence with the template mesh from \cite{ploumpis2019combining} using a landmark-guided Non-rigid Iterative Closest Point (NICP)~\cite{amberg2007optimal} registration procedure followed by Gaussian Processes, which were used to improve the deformation flexibility in syndromic cases.

        \subsection{Data augmentation with spectral interpolation.}
            \label{sec:data-aug}
            \noindent Given the limited amount of data, we propose a novel data augmentation technique operating in the spectral domain and inspired by \cite{foti2020intraoperative} and \cite{he2009learning}. Unlike random interpolations performed in the Euclidean space, the proposed augmentation produces samples where local shape features are separately combined, ultimately resulting in more diverse shapes. Since the head meshes are in dense point correspondence and share the same topology, they also have the same mesh Laplacian operator $\mathbf{L}$. The eigendecomposition of this operator, $\mathbf{L} = \mathbf{U} \bm\Lambda \mathbf{U}^T$, determines a set of orthonormal eigenvectors which can be used to compute the Fourier transform of the meshes in the dataset as $\hat{\mathbf{X}} = \mathbf{U}^T\mathbf{X}$ and the inverse Fourier transform as $\mathbf{X} = \mathbf{U}\hat{\mathbf{X}}$. Where $\mathbf{U}$ are the eigenvectors and $\mathbf{X}$ the vertex positions. With these operators, the proposed data augmentation can be defined as:
            \begin{equation}
                \label{eq:augmentation}
                    \mathbf{X}_\text{aug} =  \mathbf{X}_1 + \mathbf{U} \big[ \bm{\varrho} \; \mathbf{U}^T (\mathbf{X}_2 - \mathbf{X}_1) \big],
            \end{equation}
            with $\bm{\varrho} = \big\{\varrho_1,\ldots, \varrho_{30}, 0,\ldots, 0 \, \big| \, \varrho_{i} \sim \mathcal{N}(\mu = 0.5, \sigma = 0.5 ) \big\}$.
            Eq.~\ref{eq:augmentation} is equivalent to performing an interpolation between the spectral components of two meshes with vertices $\mathbf{X}_1$ and $\mathbf{X}_2$ respectively. Only the first $30$ spectral components, which are associated to lower frequency details of the mesh, are interpolated according to the value of different random variables $\varrho_i$. When $\varrho_i=0$, the $i$-th spectral component of $\mathbf{X}_\text{aug}$ corresponds to the $i$-th spectral component of $\mathbf{X}_1$, when $\varrho_i=1$ it corresponds to the one of $\mathbf{X}_2$.
            Since $\varrho_i$ values are drawn by Gaussian distributions centred in $0.5$ with a standard deviation of $0.5$, about $68\%$ of the interpolation values will be in the range $[0, 1]$. Therefore, some spectral components will be slightly outside the usual interpolation range, leading to more diverse augmentations.
            
            To compensate for the high class imbalance, classes with fewer training data samples are augmented more. We thus obtain $281$ training meshes per class, for a total of $1405$ meshes, which corresponds to an augmentation factor of $5$.
            
        \subsection{Swap disentangled 3D mesh variational autoencoder.}
            \begin{figure}
                \centering
                \includegraphics[width=\linewidth]{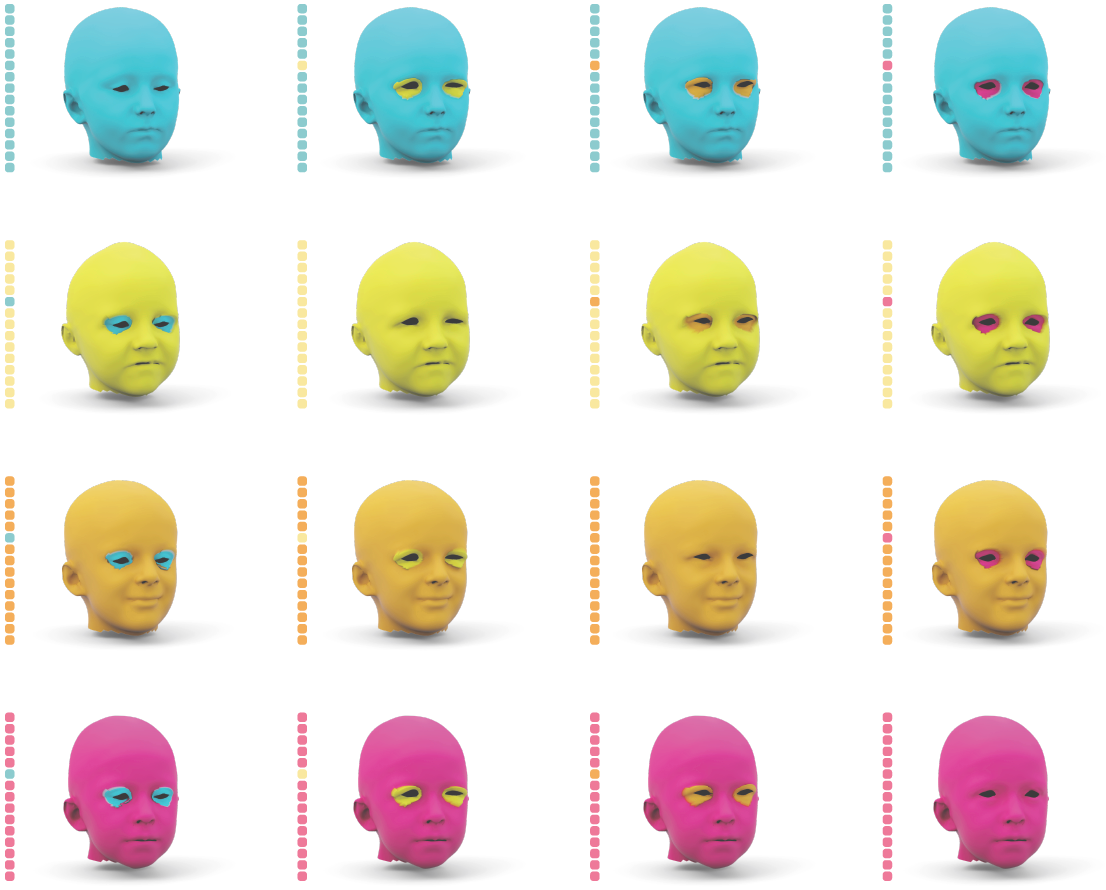}
                \caption{\normalfont  Mini-batch with swapped attributes. Each shape is showed alongside its latent representation. Both shape attributes and latent subsets are colour-coded to represent different subjects.}
                \label{fig:minibatch}
            \end{figure}
            
            \noindent A swap disentangled mesh variational autoencoder (SD-VAE)~\cite{foti20223d} was trained while encouraging disentanglement between $15$ attributes corresponding to anatomical sub-units of the head (Fig.~\ref{fig:disentangle_distribution}). As in traditional VAEs, the architecture is built with a pair of neural networks. The encoder network $E$ encodes input meshes as Gaussian distributions over the latent space and the generator $G$ generates new meshes from latent vectors. SD-VAE~\footnote{Original implementation: \href{https://github.com/simofoti/3DVAE-SwapDisentangled}{github.com/simofoti/3DVAE-SwapDisentangled}} differs from a traditional mesh VAE only in the training procedure, where data are mini-batched by swapping head attributes across different subjects to create known shape differences and similarities that can be leveraged by a latent consistency loss.
            In particular, every mini-batch can be thought of as a matrix where the elements on the diagonal are head shapes from the training set, while the other elements are obtained by swapping head attributes. At every iteration, an arbitrary attribute is selected and swapped across columns.  After the swapping, each column contains different heads with the same attribute, while each row of the matrix contains the same head shape with different attributes (Fig.~\ref{fig:minibatch}). Attributes are manually defined by segmenting the mesh template. This segmentation is inherited by all shapes as they are in dense point correspondence with the template (Sec.~\ref{sec:data}).
            The latent consistency loss aims to enforce the same differences and similarities in the latent representations of the shapes in each mini-batch. In practice, the latent representation is partitioned in as many subsets as shape attributes. Then, the subsets of latent variables obtained from the shapes in the matrix-shaped mini-batch are compared between each others (Fig.~\ref{fig:minibatch}). The difference between subsets controlling the attribute swapped at the current iteration, and obtained encoding shapes in a column, are forced to be smaller by a margin than the differences of the corresponding subsets obtained encoding shapes in a row. Similarly, the differences of all the subsets controlling all the attributes that are not swapped, and obtained from rows, are forced to be smaller by a margin then the column differences in those subsets. 
            The other losses used during training are: a mean-squared-errors-based reconstruction loss encouraging the vertices of the output mesh to be as close as possible to the vertices of the input mesh, a Kullback–Leibler (KL) divergence making the encoded distributions standard Gaussians, and a Laplacian regularisation term smoothing the output meshes to avoid surface discontinuities caused by the feature swapping. Like in the original implementation of SD-VAE~\cite{foti20223d}, $E$ and $G$ have $4$ spiral convolutional layers~\cite{gong2019spiralnet++} interleaved with pooling operators defined during a quadric-sampling procedure~\cite{gong2019spiralnet++, ranjan2018generating}. All convolutional layers have $32$ features except the last convolutional layer of $E$ and the first of $G$, which have $64$ features. Spirals have a length of $9$ and dilation of $1$, while the sampling factor is set to $4$. The last convolutional layer of $E$ is followed by two linear layers predicting the mean and standard deviation vectors of the latent distributions. Latent vectors sampled from the latent distribution are processed by another linear layer and then by all the convolutional layers in $G$. Latent vectors are made of $75$ variables grouped in $15$ subsets of $5$ variables. Each subset controls a different head attribute. Data were divided in three subsets using a stratified split. $80\%$ of data were used during training, $10\%$ during validation, and the remaining $10\%$ during testing. The network was trained with the Adam optimiser~\cite{kingma2014adam} for $600$ epochs using a batch size of $16$ and a learning rate of $1e^{-4}$. The Laplacian, KL, and latent consistency loss weights were set to $\alpha=0.1$, $\beta=1e^{-4}$, and $\kappa=0.5$ respectively. The margin values in the latent consistency loss were set to $\eta_1 = \eta_2 = 0.5$. 
            The above mentioned parameters were the same as the original implementation of SD-VAE, except for the number of epochs, the Laplacian regulariser weight and the latent consistency weight. Both weights were determined with a grid search reducing and increasing the original parameters by one order of magnitude. More epochs were required as a consequence of the smaller dataset.
        
        \subsection{Manifold visualisation.} 
            \label{sec:manifold-methods}

            \begin{figure}[t]
                \centering
                \includegraphics[width=\linewidth]{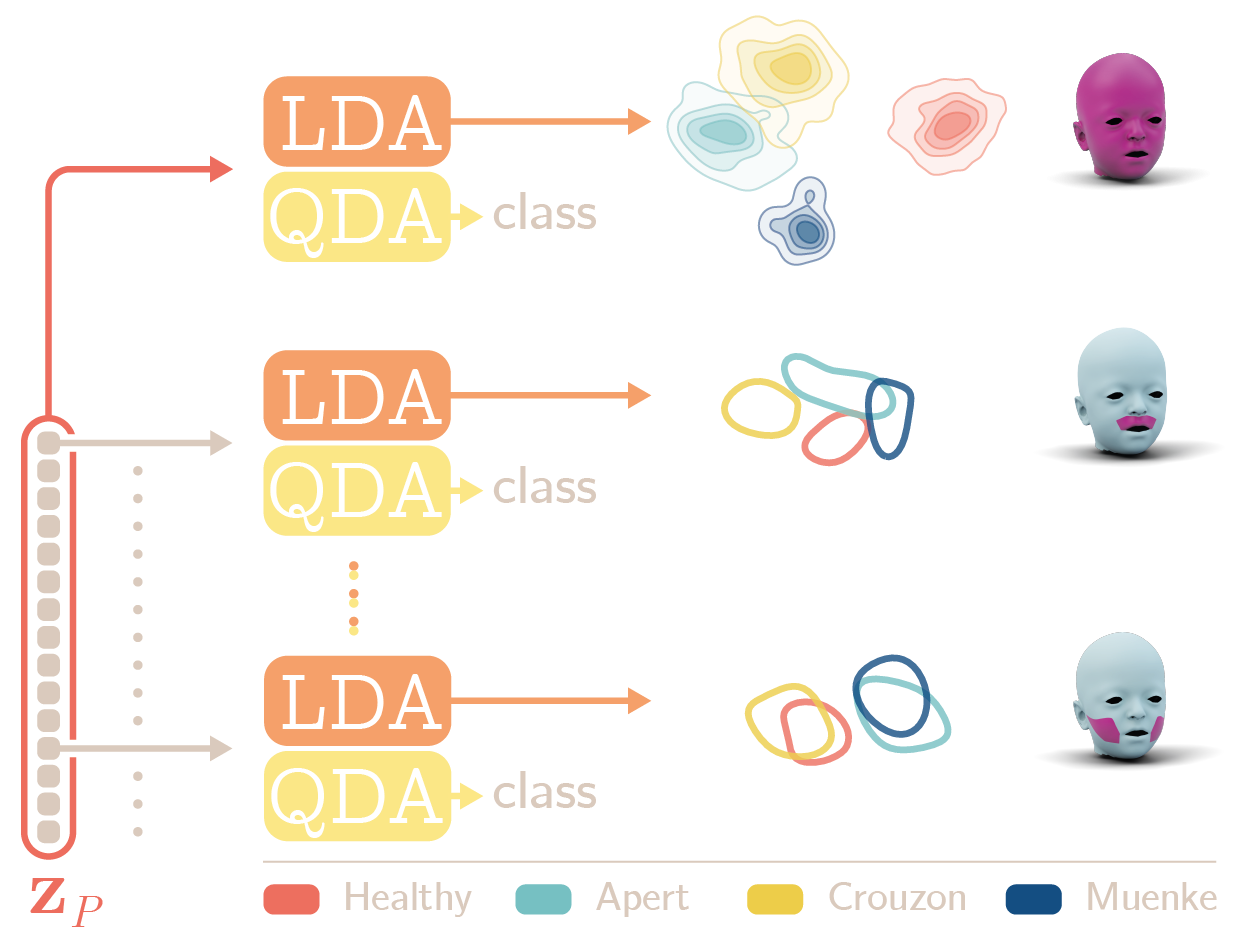}
                \caption{\normalfont Manifold visualisation and syndrome classification. Latent vectors obtained encoding shapes with SD-VAE, $\mathbf{z}_P$, can be processed as a whole or in subsets of variables. LDA and QDA models are thus created for each latent subset and for the the whole latent. LDA models are used for dimensionality reduction and manifold visualisation, while QDA models for classification. }
                \label{fig:lda-qda}
            \end{figure}
            
            \noindent Once trained, the encoder of SD-VAE predicts the latent distribution associated to each head shape. 
            We consider the latent vectors corresponding to the means of the predicted latent distributions.
            To visualise these high-dimensional vectors we rely on the dimensionality reduction properties of the Linear Discriminant Analysis (LDA) technique.
            LDA is a supervised method that maximises class separability. Its main advantages over other dimensionality reduction techniques, such as Principal Component Analysis (PCA) or t-distributed Stochastic Neighbour Embedding (t-SNE), are that (\emph{i}) LDA computes a linear transformation that can be used to project new data, and (\emph{ii}) it highlights whether latent vectors corresponding to different classes are linearly separable.
            
            We use LDA to project the 75-dimensional latent vectors onto the two axes, maximising the distance between the means of the different classes and minimising the inter-class variances. Not only we use LDA to reduce the dimensionality of the entire latent vectors, but also to reduce and visualise the 5-dimensional subsets of the latent vectors controlling the different anatomical sub-units of the head (see Fig.~\ref{fig:lda-qda}). Visualising these manifolds gives us a better understanding of the diagnostic capacity of the model and provides insights on the most statistically important anatomical sub-units to diagnose the different syndromes. In addition, we can project the latent vector and the 5-dimensional subsets of new patients in the manifold visualisations to aid both diagnosis and surgical planning. The LDA models are fitted with the Singular Value Decomposition solver using a tolerance threshold of $1e^{-4}$ and no shrinkage.

        \subsection{Syndrome classification.}
            \label{sec:classification-methods}
            \noindent Even though LDA is also a classification technique, it assumes that data from different classes follow a normal distribution where the covariance matrix is the same across classes. To relax this assumption, we rely on QDA, a generalisation of LDA which computes different covariance matrices for different classes. Like for the manifold visualisations, we operate on the latent vectors (and their subsets of variables) corresponding to the means of the latent distributions predicted by SD-VAE (see Fig.~\ref{fig:lda-qda}). The QDA models are fitted with an absolute tolerance threshold of $1e^{-4}$ and no per-class covariance regularisation.

        \subsection{Evaluating the effects of different surgical approaches using latent interpolation.} 
            \label{sec:surgical-planning}

            \begin{figure}[ht]
                \centering
                \includegraphics[width=.5\linewidth]{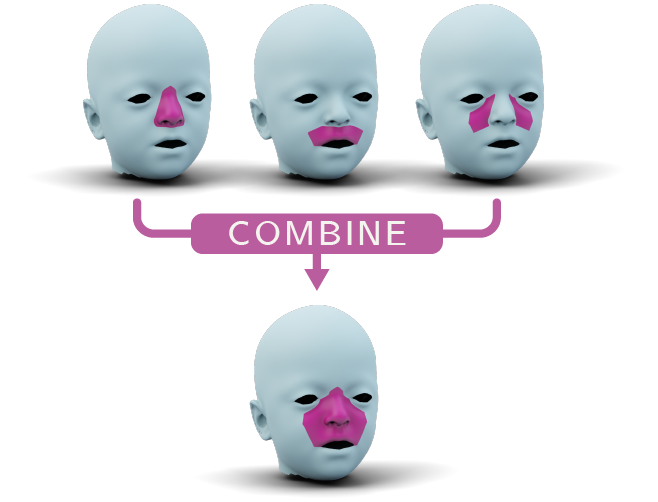}
                \caption{\normalfont Example of surgical region definition. In this case, the nose is combined with the upper lip and the nasolabial attributes, thus defining the regions affected by the Le Fort II surgical procedure}
                \label{fig:combine}
            \end{figure}
            
            \noindent One of the main advantages of VAEs over traditional autoencoders, is their ability to generate new data. SD-VAE inherits this capability, which we leverage to aid surgical planning. Surgical intervention for the correction of syndromic craniosynostosis has both functional and aesthetic purposes. We hypothesise that a surgical procedure capable of moving the shape of the patient's head closer to the distribution of healthy shapes will improve their appearance and have the potential to ameliorate functional impairments. 
            Therefore, we can assert that it is desirable to move the latent representation of the patient's head ($\mathbf{z}_P$) towards the centre of the latent distribution of healthy subjects ($\mathbf{z}^H_\mu$). If we linearly interpolate latent vectors between $\mathbf{z}_P$ and $\mathbf{z}^H_\mu$, we can generate their corresponding shapes and simulate the ideal transition towards an healthy head. However, it is worth noting that obtaining the mean shape of healthy subjects is not always desirable if we want to preserve some of the shape properties characterising the identity of the syndromic patient. In addition, standard surgical procedures do not involve all the anatomical sub-units of the head and a latent interpolation over the entire latent is not informative. To overcome the first issue, instead of considering $\mathbf{z}^H_\mu$ the only ideal outcome, we consider three additional target latent vectors to control where to stop the interpolation: $\mathbf{z}^H_{1\sigma}$, $\mathbf{z}^H_{2\sigma}$, $\mathbf{z}^H_{3\sigma}$. These vectors lie on the line passing through $\mathbf{z}_P$ and $\mathbf{z}^H_\mu$, at $1$, $2$, and $3$ standard deviations from $\mathbf{z}^H_\mu$, where the standard deviation is computed over the latent distribution of healthy subjects. To overcome the second issue, we rely on the disentangled latent representation of SD-VAE, performing the linear interpolation only on the latent variables controlling regions affected by the surgical procedures.
            These regions are defined by combining anatomical sub-units disentangled by SD-VAE (Fig.~\ref{fig:combine}). We qualitatively evaluate the effects of the procedure-specific latent interpolations embedding the latent vectors with LDA and comparing their position with respect to the healthy distribution. In addition, we quantitatively evaluate which procedure brings the encoded shape closer to $\mathbf{z}^H_\mu$. This is assessed by measuring the Euclidean distances between $\mathbf{z}^H_\mu$ and the latent vectors obtained with the procedure-specific interpolation stopped at $\mathbf{z}^H_\mu$, $\mathbf{z}^H_{1\sigma}$, $\mathbf{z}^H_{2\sigma}$, and $\mathbf{z}^H_{3\sigma}$. 

            Note that not only surgeons can control the local interpolations by selecting different stopping points for each attribute and latent variable, but they can also choose other latent vectors belonging to the healthy distribution as suitable targets. Both the attribute-specific stopping vectors and the alternative targets could be manually selected by the surgeon via a graphical user interface.  In addition, if a target shape is available, it can be encoded with our model and its latent representation can be used as the target of the local interpolation. The target shape may be obtained by manually sculpting the patient's head-shape during a pre-operative consultation, by selecting it from a library of predefined shapes, or by scanning a new healthy subject.

    \section{Results}

        \subsection{Data augmentation.}
\begin{figure*}
                \centering
                \includegraphics[width=\textwidth]{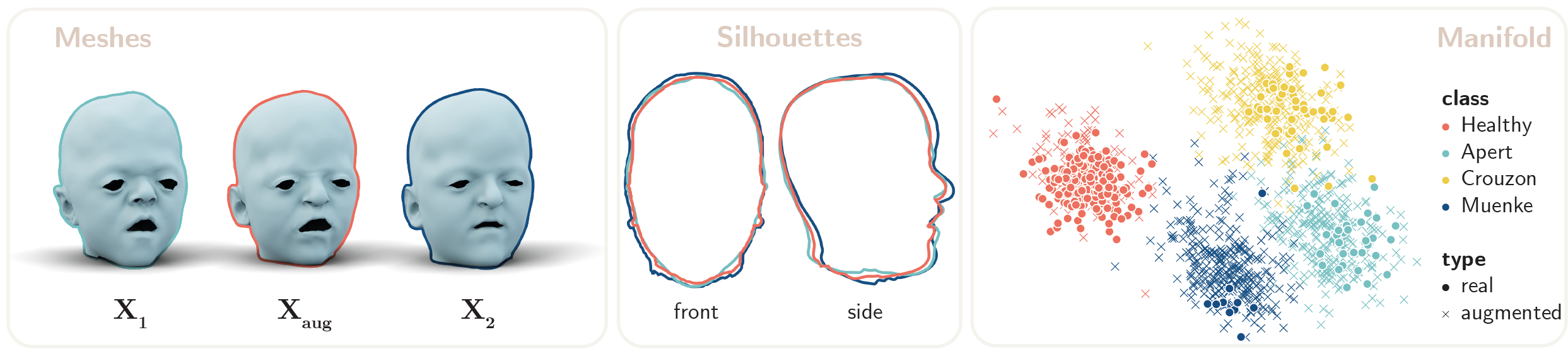} 
                \caption{\normalfont Evaluation of the data augmentation. \textit{Left:} rendering of the mesh whose vertices ($\mathbf{X}_\text{aug}$) were obtained with the proposed data augmentation technique from the two real Apert meshes with vertices $\mathbf{X}_1$ and $\mathbf{X}_2$. \textit{Centre:} silhouettes of the real and augment meshes in the front and side view. The silhouettes are colour-coded like their corresponding meshes. \textit{Right:} manifold visualisation of SD-VAE trained on real data only.}
                \label{fig:augmentation}
            \end{figure*}

            \noindent The dataset introduced by \cite{o2021craniofacial} contains only a limited amount of data and it exhibits a significant class imbalance. Our novel data augmentation technique (see Sec.~\ref{sec:data-aug}) overcomes these limitations, augmenting the dataset size and balancing the number of samples in each class. In order to maintain a consistent class labelling, new meshes are created by randomly interpolating the spectral components of two real meshes sampled from the same class. In addition, observing how age influences the spectra of the real data (see Supplementary Materials), we notice the emergence of age-related clusters. For this reason, the two real meshes used to create new data are always sampled from the same age group, which can be either $[0,4)$ or $[4,20]$ years old. Since no sex-related cluster was noticed, the meshes can represent both male and female subjects.
            In Fig.~\ref{fig:augmentation}~(\textit{left}), we depict the augmented mesh with vertex positions $\mathbf{X}_\text{aug}$ and the two real Apert meshes used for the augmentation, where $\mathbf{X}_1$ is a $6$-year-old female and $\mathbf{X}_2$ a $12$-year-old female. Observing these meshes and their corresponding silhouettes in  Fig.~\ref{fig:augmentation}~(\textit{centre}), we notice that, not only does the augmented sample maintain the morphological features of Apert patients, but it also plausibly combines features from the real patients. 
            
            To prove that augmented data are properly labelled and are suitable to train our model, we visualise the latent manifold by embedding the latent vectors obtained encoding real and augmented shapes with SD-VAE trained on all the real data. Even though the augmented shapes were not used to train SD-VAE, we observe that they cluster together near the embeddings of real shapes, and they create a richer and more well rounded latent space. Therefore, we consider the augmented data to be statistically suitable for training the SD-VAE, the QDA classifiers, and the LDA models used for the manifold visualisation. 

        \subsection{Model and latent disentanglement evaluation.}
            
            \noindent SD-VAE reports a reconstruction error of $\SI{1.27 \pm 0.34}{\mm}$. This value was computed as the mean errors between all the meshes from the test set and their reconstructions. Errors are always calculated with the mean Euclidean distance between the vertices of two meshes. We then evaluated the generative performances of SD-VAE in terms of diversity, which is computed as the mean errors between pairs of $10,000$ randomly generated meshes. A large diversity value indicates that the model can generate a wider variety of output shapes. SD-VAE reports a diversity of $\SI{5.94}{\mm}$. When SD-VAE is trained only with real data, reconstruction errors increases to $\SI{1.72 \pm 0.69}{\mm}$ and the diversity decreases to $\SI{4.47}{\mm}$, further motivating the introduction of our data augmentation.
            
\begin{figure*}[h]
                \centering
                \includegraphics[width=\textwidth]{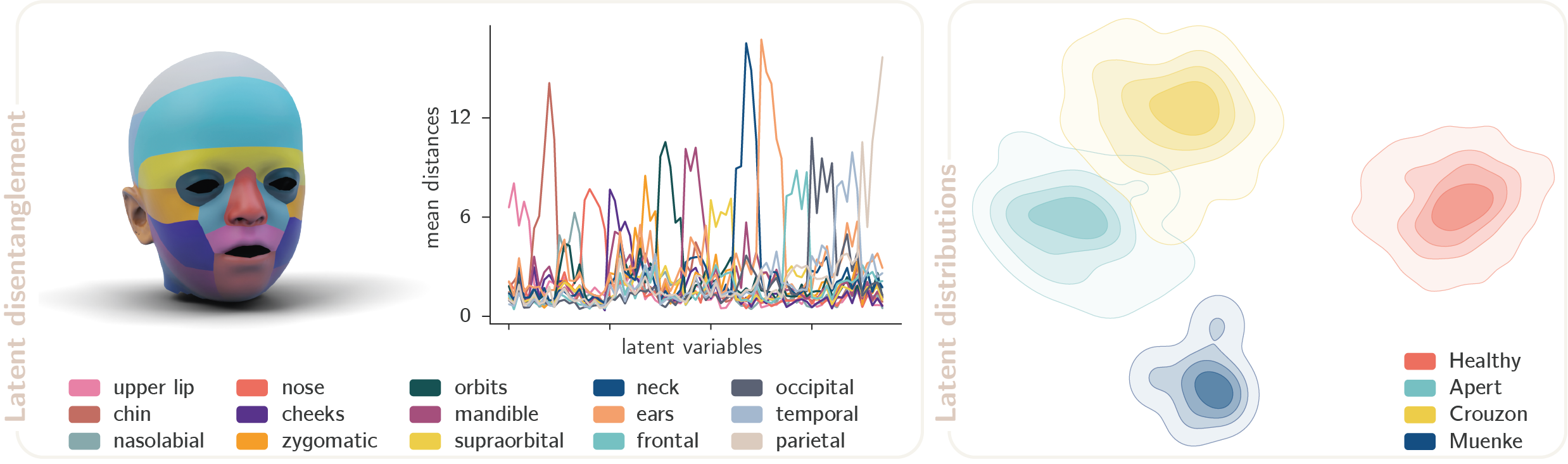} 
                \caption{\normalfont Latent evaluation. \textit{Left:} visual representation of the $15$ anatomical attributes for which we seek to obtain a disentangled latent representation and the effects of traversing each latent variable (abscissas). The mean distance between a mesh generated with a given variable at its minimum and the one generated with the variable at its maximum is reported for each attribute (ordinates). For each latent variable, we expect a high mean distance in one single attribute and low values for all the others. \textit{Right:} manifold visualisation of SD-VAE trained on both real and augmented data. The distributions of the latent embeddings are colour-coded according to the different syndromes.}
                \label{fig:disentangle_distribution}
            \end{figure*}

            A disentangled latent representation where the different anatomical shape attributes can be separately analysed and generated is essential for the local analysis of the different syndromes and for procedure-specific planning.
            Although multiple metrics have been proposed to quantitatively evaluate latent disentanglement, they require labelled data or control over the generative factors~\cite{foti20223d}. Since these information are not available, we evaluate latent disentanglement by measuring the average vertex displacement caused in each anatomical sub-unit of the generated meshes while traversing each latent variable~\cite{foti20223d}. Plotting the average displacements for each attribute in isolation offers an intuitive approach to evaluate disentanglement. Effective disentanglement is achieved when manipulating a single variable leads to significant displacements for one attribute while maintaining low mean distances for the remaining attributes. As we can observe in Fig.~\ref{fig:disentangle_distribution}~(\textit{left}), all the head attributes are properly disentangled. In fact, the $15$ subsets of latent variables are related to only one attribute each. When it is apparent that traversing a latent variable does in fact create displacements in other anatomical sub-units, we notice that these displacements are generally small and in neighbouring units. These displacements are necessary to preserve the surface continuity between neighbouring units.

        \subsection{Manifold visualisation.}
            \noindent We embed high dimensional latent vectors in a two-dimensional space applying the linear transformation computed via Linear Discriminant Analysis (see Sec.~\ref{sec:manifold-methods}). The distributions of the embedded latent vectors are displayed in Fig.~\ref{fig:disentangle_distribution}, using a different colour for each syndrome. All classes appear to be linearly separable with no significant overlap between the distributions. This corroborates the previous findings of \cite{o2021craniofacial}, proving that the different syndromes can be easily identified from latent vectors encoding a 3D mesh of the head. 
            
\begin{figure*}
                \centering
                \includegraphics[width=\textwidth]{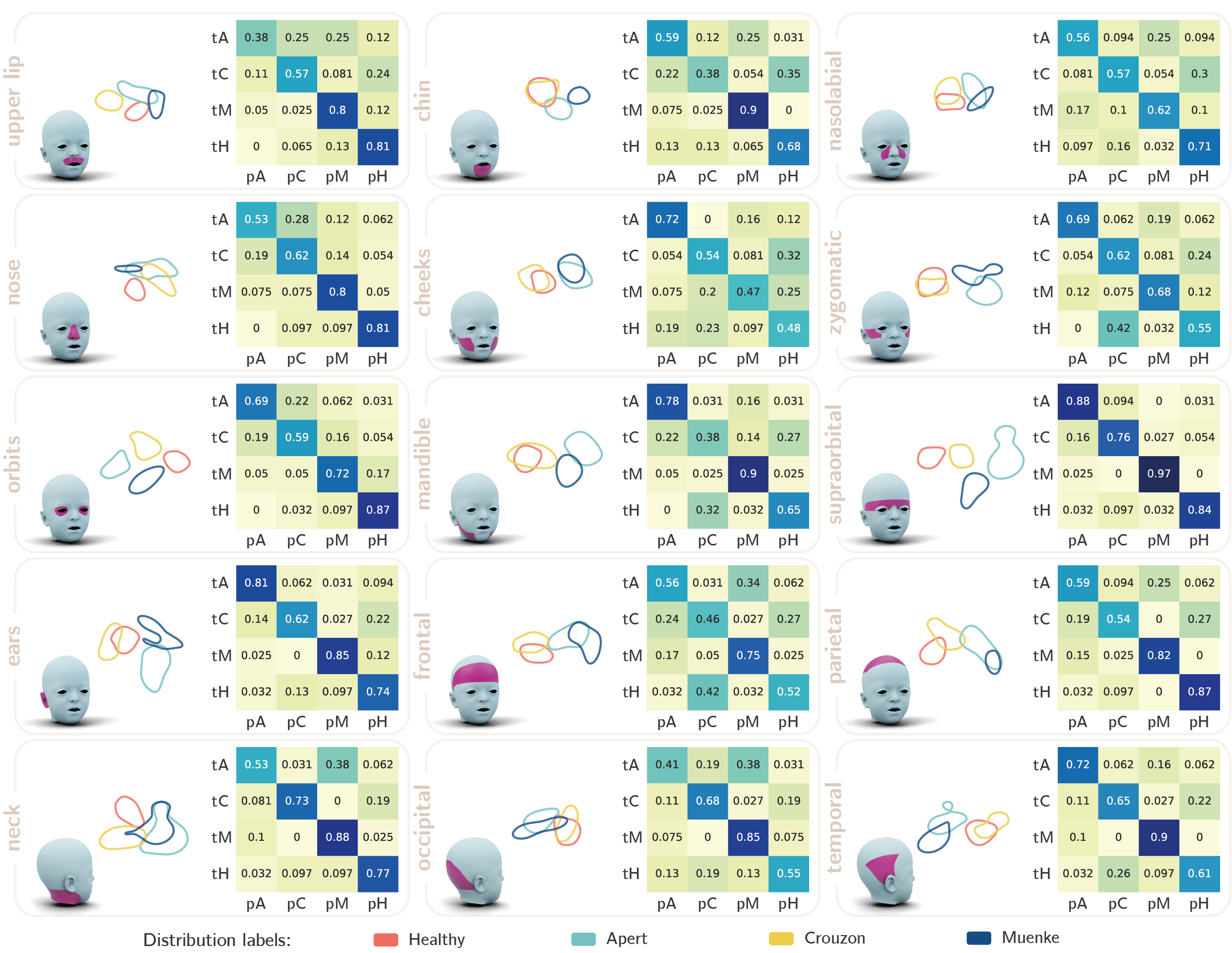} 
                \caption{\normalfont Attribute-specific evaluation. For each attribute we depict the corresponding anatomical sub-unit rendered on the template head (\textit{left}), the manifold visualisation of the per-class distributions (\textit{centre}), and the confusion matrix reporting the classification results on the test data of the attribute-specific QDA classifier (\textit{right}). The iso-contour lines representing the different distributions are drawn at one standard deviation. The rows of the confusion matrices correspond to the true labels "tL", while columns to the predicted labels "pL", where L = A for Apert, C for Crouzon, M for Muenke, and H for Healthy.}
                \label{fig:attributes}
            \end{figure*}

            Thanks to the more interpretable, structured, and disentangled latent representation of SD-VAE, we can also embed the subsets of latent variables controlling local shape properties of different anatomical sub-units of the head. As for the complete latent vectors, their subsets are also embedded with LDA.
            Observing Fig.~\ref{fig:attributes}, we notice that different facial features hold greater influence over the syndromic phenotype. A smaller overlap between pairs of distributions indicate a higher class separability, which translates to an increased importance of the given attribute during classification. The orbits, the upper lip, and the supraorbital regions have well separated distributions. Therefore, they appear to be particularly important when identifying healthy subjects and diagnosing patients with syndromic craniosynostosis. The distribution of attributes such as the nose and the neck have no overlap between the healthy and syndromic populations. However, there is significant overlap amongst the three craniofacial syndromes. This shows how the shape of these attributes can be used to diagnose syndromic patients, but are less useful in identifying the correct syndrome. Interestingly, the iso-contour lines for both the healthy and Crouzon populations, as well as those representing the Muenke and Apert distributions, are often overlapped. This demonstrates that there are similarities in shape between these classes in certain anatomical sub-units.  

        \subsection{Syndrome classification.}
            \noindent We perform multi-class classification on the entire latent vectors and on their subsets which are controlling the different attributes. As detailed in Sec.~\ref{sec:classification-methods}, multiple Quadratic Discriminant Analysis (QDA) models are fitted: one to perform classification on the whole latent and $15$ models to perform classification on the latent subsets. Similarly to \cite{o2021craniofacial}, the classification of the entire latent vectors encoded from the test shapes yields a $100\%$ accuracy with QDA. The average precision, recall, and F1-score are also $100\%$. This result further demonstrates how it is possible to classify healthy, Apert, Crouzon and Muenke patients from the shape of their head with great accuracy. However, as we noticed by analysing the manifold distributions of the different attributes, the classification is not equally trivial if single anatomical sub-units are examined. We report the performance of the attribute-specific QDA classifiers as confusion matrices (Fig.~\ref{fig:attributes}). Similarly to the overlaps between the manifold distributions, these matrices can be used to evaluate the importance of each attribute in the diagnosis of different syndromes. The diagonal elements quantify the number of latent subsets accurately classified. In addition, the other values tell us where the classifier failed and how often. 
            
            Note that when SD-VAE is trained without data augmentation, the classification accuracy of QDA on the entire latent falls from $100\%$ to $71.88\%$. Similarly, average precision, recall, and F1-score fall from $100\%$ with augmentation to $51.66\%$, $71.88\%$, and $60.11\%$ without augmentation. This proves once again the effectiveness and the importance of the proposed data augmentation.

            \begin{figure*}[ht]
                \centering
                \includegraphics[width=\textwidth]{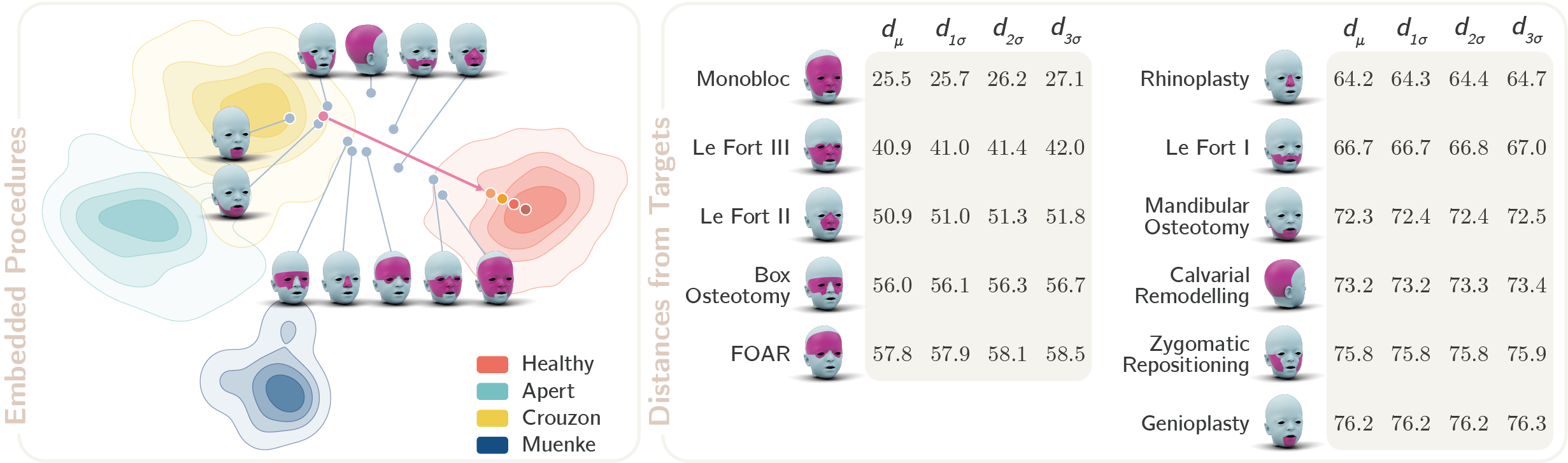} 
                \caption{\normalfont Comparison between the surgical procedures that could be performed on a Crouzon patient. The area of influence of each procedure is defined by combining the attributes disentangled by SD-VAE.
                \textit{Left:} embedding of all the surgical procedures. The latent representation of the patient's head ($\textbf{z}_P$) is embedded into the two-dimensional space as a pink dot. The pink arrow represents the ideal interpolation trajectory towards the distribution of healthy subjects when all latent variables are interpolated. The interpolation can be stopped at four points: $\mathbf{z}^H_{3\sigma}$, $\mathbf{z}^H_{2\sigma}$, $\mathbf{z}^H_{1\sigma}$, and $\mathbf{z}^H_\mu$. Their embeddings are represented by the four orange dots between the point of the arrow and the mean of the healthy distribution. Blue dots represent the embedding of the procedure-specific latent interpolations when the latent variables associated to the anatomical sub-units affected by the procedure are interpolated until their $\mathbf{z}^H_\mu$ value. \textit{Right:} latent distances between $\mathbf{z}^H_\mu$ and the procedure-specific interpolations stopped at $\mathbf{z}^H_\mu$ ($d_\mu$), $\mathbf{z}^H_{1\sigma}$ ($d_{1\sigma}$), $\mathbf{z}^H_{2\sigma}$ ($d_{2\sigma}$), and $\mathbf{z}^H_{3\sigma}$ ($d_{3\sigma}$). Procedures are ranked from most to least effective.}
                \label{fig:procedures}
            \end{figure*}
            
        \subsection{Evaluating the effects of different surgical approaches.}

            \noindent As mentioned in Sec.~\ref{sec:surgical-planning}, we hypothesise that a surgical procedure capable of bringing the shape of the patient's head closer to the centre of the healthy distribution of Fig.~\ref{fig:disentangle_distribution}, will improve the aesthetic appearance of the patient and also potentially address functional impairments. We start by encoding the patient's head with SD-VAE to obtain its latent representation $\textbf{z}_P$. Then, we identify the centre of the healthy latent distribution $\mathbf{z}^H_\mu$ as well as three additional latent vectors where the interpolation can be stopped ($\mathbf{z}^H_{1\sigma}$, $\mathbf{z}^H_{2\sigma}$, $\mathbf{z}^H_{3\sigma}$). We embed and visualise all these vectors on top of the manifold visualisation (Fig.~\ref{fig:procedures}, \textit{Left}) and, as expected, the Crouzon patient is projected on top of the Crouzon distribution. 
            
\begin{figure*}[ht]
                \centering
                \includegraphics[width=\textwidth]{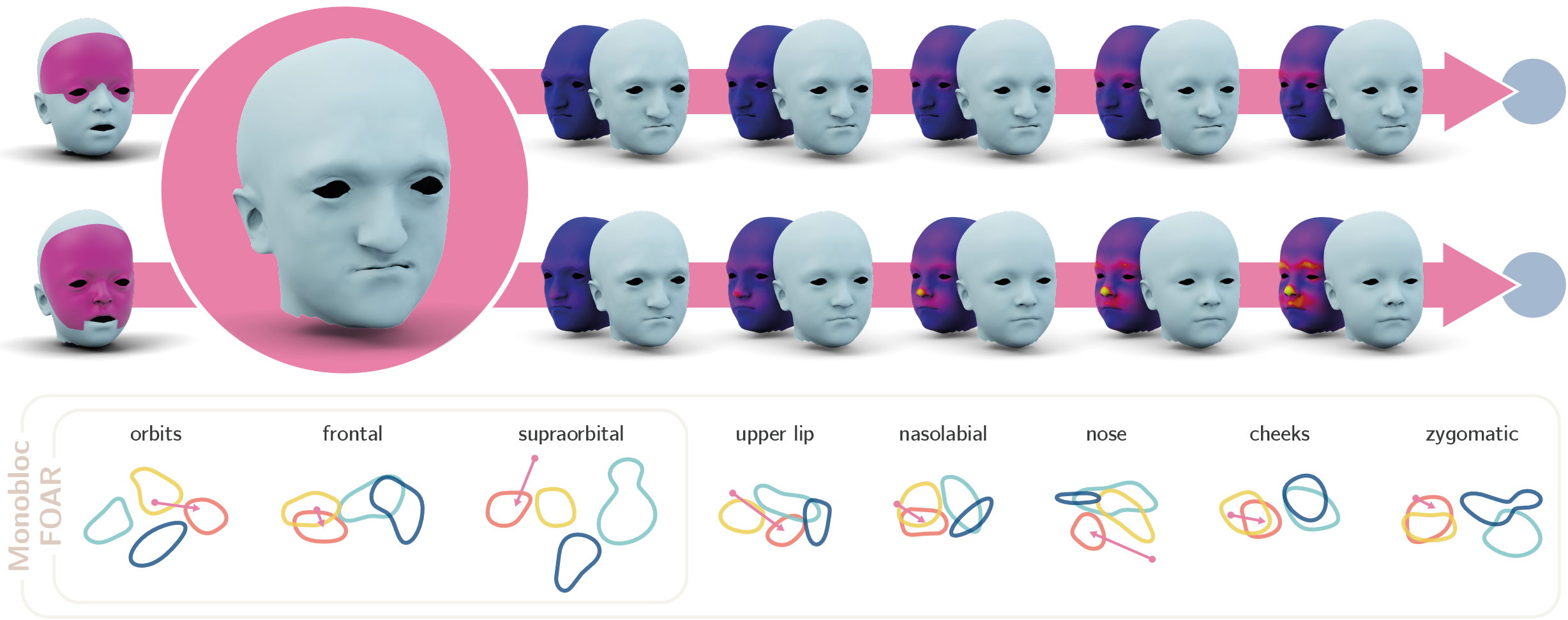} 
                \caption{\normalfont Monobloc and FOAR latent interpolations for a Crouzon patient. The patient is the same of Fig.~\ref{fig:procedures}. \textit{Top:} the patient's head is represented in the pink circle. On its left, the anatomical sub-units affected by the procedures are highlighted. On the right of the pink circle, the shapes generated during the procedure-specific interpolation are rendered alongside a displacement map showing the magnitude of the displacements (blue corresponds to no displacement and yellow to displacements of more than $10$ mm). Both interpolations stop at the blue points represented in Fig.~\ref{fig:procedures}. Note that the figure represents the interpolation from the original shape to the target in order to demonstrate the locality of the shape alterations as well as the smoothness of the transition. However, the surgeon can also select intermediate shapes as desired surgical outcome.  \textit{Bottom:} embeddings of the latent interpolations for the different subsets of latent variables controlling the anatomical sub-units affected by the two procedures. Although all the affected anatomical sub-units are interpolated until they reach the target, surgeons can control the local interpolations by selecting different stopping points along the predefined trajectory of each attribute. }
                \label{fig:p_interpolation}
            \end{figure*}

            Owing to the technical limitations of surgery, surgical procedures tend to focus on anatomical sub-units, exacting change at a local level. Thanks to the more interpretable, structured, and disentangled latent representation of SD-VAE, we can analyse and simulate the effects caused by the most common craniofacial procedures. For each procedure we group the subsets of latent variables controlling the shape attributes affected by the procedure and we interpolate them between their original $\textbf{z}_P$ value and their target $\mathbf{z}^H_\mu$, $\mathbf{z}^H_{1\sigma}$, $\mathbf{z}^H_{2\sigma}$, and $\mathbf{z}^H_{3\sigma}$ values. The latent vectors obtained for every procedure when the latent interpolation is stopped at $\mathbf{z}^H_\mu$ are embedded and visualised in Fig.~\ref{fig:procedures}~\textit{left} as blue dots. This representation provides an immediate indication of which procedures bring the shape of the patient's head closer to the centre of the healthy distribution. However, when a latent vector is embedded in a two dimensional space, some information is lost. In fact, if like in Fig.~\ref{fig:procedures}~\textit{right} we measure the distance between $\mathbf{z}^H_\mu$ and the latent vectors obtained stopping the procedure-specific interpolations at $\mathbf{z}^H_\mu$ ($d_\mu$), $\mathbf{z}^H_{1\sigma}$ ($d_{1\sigma}$), $\mathbf{z}^H_{2\sigma}$ ($d_{2\sigma}$), and $\mathbf{z}^H_{3\sigma}$ ($d_{3\sigma}$), we notice that some procedures are closer to the target than they appear when embedded (e.g., box and mandibular osteotomies). In addition, stopping the procedure-specific latent interpolations before reaching the centre of the healthy distribution does not significantly affect the metrics. It is thus sufficient to observe $d_\mu$ to identify the procedure more capable of ameliorating the patient's head shape. In Fig.~\ref{fig:p_interpolation}~\textit{top} we show the head shapes generated during the monobloc and FOAR interpolations of the same Crouzon patient analysed in Fig.~\ref{fig:procedures}. These shapes and their associated displacement maps are obtained at five equidistant locations along the interpolation trajectory and can be examined to understand how the shape of the different anatomical sub-units could be changed. Note how the procedure-specific interpolation affects only the anatomical sub-units, correlating with the regions contributing the most to the patient's syndromic appearance. 
            
            Observing the effects of the latent interpolations on the attribute-specific latent distributions (Fig.~\ref{fig:p_interpolation}~\textit{bottom}), we notice that this patient deviates from the healthy distribution especially in the orbits, supraorbital, upper lip, and nose attributes. These are also the attributes whose shape is mostly affected during the interpolations depicted in Fig.~\ref{fig:p_interpolation}~\textit{top}. Therefore, similarly to Fig.~\ref{fig:p_interpolation}~\textit{bottom}, projecting a patient's head in the attribute-specific manifold visualisation of Fig.~\ref{fig:attributes}, provides important information on which attributes are causing the patient to be perceived as syndromic. Not only this is useful in aiding selection of the most appropriate surgical procedure, but it can also help interpret complex cases where machine learning outperforms expert diagnosis~\cite{o2021craniofacial} (see Supplementary Materials).

    \section{Discussion}

        \noindent SD-VAE is a novel methodology which allows for the disentanglement of anatomical sub-units and objective classification of their influence on the overriding phenotype. Not only does this allow for multi-class diagnosis with high sensitivity and specificity but also provides insight into how we can improve treatment planning and surgical management of deformity. 
        
        As shown in Fig.~\ref{fig:disentangle_distribution} the latent disentanglement of the fifteen facial features was achieved with a high level of accuracy. Perhaps most interestingly, Fig.~\ref{fig:attributes} elucidates the regions which are most key to syndromic identity. For example, frontal and periorbital regions are shown to play a significant role in defining the appearance in Muenke, Crouzon and Apert syndrome. This is reflected in the surgical approach to these patients, with fronto-orbital remodelling and midfacial osteotomies traditionally being a mainstay of treatment. Where regions have no overlap in iso-contour lines, it highlights the clear morphological differences in features between each syndrome. 
        
        There is a significant psychological burden experienced by those with congenital craniofacial deformities, be it as result of difficulties at school, through preconceptions of ability or in reduced self-confidence~\cite{endriga1999psychological}. As a result, craniofacial surgery often has not only a functional but also an aesthetic component aiming at normalising appearance. Certainly, there are problems with using a normal head shape as a reference point for guiding surgical planning. Even though the introduction of multiple stopping points during the latent interpolation mitigates this problem, an AI perception of normality or indeed beauty holds an ethical quandary and inherent bias from the dataset on which the model is fabricated \cite{liang2021artificial,koimizu2019machine,patcas2019applying,fayemi2018racial}. This is not to say that technology is to replace the human dimension of healthcare or the personal component of plastic surgery \cite{bouletreau2019artificial,jarvis2020artificial}. Instead, the suggestion would be to use this as a reference and starting point for patient education, outcome assessment and streamlining of surgical planning. 
        
        Although virtual surgical planning (VSP) has been shown to optimise surgery, it preserves an element of subjectivity which demands a high level of clinical acumen to translate into good outcomes~\cite{efanov2018virtual}. In order to automate VSP, it is necessary to control the regional anatomy and understand how this relates to normal variation. For example, by dividing the face into anatomical sub-units, we can objectively understand how much a jaw discrepancy is affecting a patient’s appearance. Taking this information, we can then understand how correcting this deformity through surgery would move both this region and the patient’s head shape towards normality.
        Regardless, the missing link towards accurate prediction of outcomes in VSP remains the open problem of depicting the non-linear relationship between the craniofacial skeleton and the position of the overlying soft tissues~\cite{madsen2018probabilistic,olivetti20193d,lubkoll2014optimal}. Although acquiring the necessary volume of pre- and post-operative surgical data may be challenging, SD-VAE has the potential to help us better understand this complex relationship. 
        
        When looking at rare diseases, the large data requirements of AI systems are hindered by a paucity of data. The need for data augmentation through spectral interpolation, whilst a solution on one hand, comes with a caveat. Although the process of augmentation is robust, reduced reconstructive errors, and significantly increased the classification accuracy, real world data would be preferred. Unfortunately, without cloud-based platforms to integrate data collection from large craniofacial centres internationally, augmentation will always be required. This approach is complicated by the sensitive nature of patient identifiable meshes and the need for standardised consent and data-use agreements~\cite{kohli2018ethics, long2017artificial}. In spite of this, a collaborative approach to data collection would ensure that any derived model truly represents the patient population, includes a diverse range of ethnicities and reduce inherent bias~\cite{liang2021artificial,jarvis2020artificial,fayemi2018racial, hashimoto2018artificial}. A collaborative approach will also generally benefit data collection of rare craniofacial syndromes, thus potentially enabling the creation of models capable of operating on more syndromes.
        
        In future work we will aim to further improve SD-VAE by disentangling the patient's age as a separate latent variable capable of conditioning the generation process. This will be essential to make sure that the latent interpolation can be performed towards the centre of the distribution of healthy subjects with the same age range of the patient. In addition, this will make it possible to predict how the shape of the patient's head will change with growth, as well as the morphological outcomes of specific surgical procedures when undertaken at different time points. This would provide insight into the optimal timing for surgical intervention, which would be of great clinical value. Again, highlighting the need for widened data collection, only a limited number of subjects are currently available for certain age-groups and syndromes. As a result, even our data augmentation technique, when applied to pairs of subjects with approximately the same age, could not be leveraged to synthesise the data required to guide the disentanglement of age. Another potential future direction involves the exploration of geodesic latent trajectories capable of following the geometry of the latent manifold~\cite{arvanitidis2017maximum, arvanitidis2018latent}. As the geodesic shortest paths are determined by prioritising regions with more data, we believe that a widened data collection for each syndrome must be performed first.
        
        In summary, the ability of SD-VAE to disentangle facial features and determine how each region contributes to the overall craniofacial appearance has great translation potential. Avenues for future research and integration into clinical practice lie in the automation of virtual surgical planning and objective evaluation of outcomes.

    \section*{Acknowledgments}
        \noindent This research was funded in part by the Wellcome Trust [203145Z/16/Z]. For the purpose of Open Access, the author has applied a CC BY public copyright licence to any Author Accepted Manuscript version arising from this submission. This work was also supported by the Great Ormond Street Hospital Charity Clinical Research Starter Grant (award n. 17DD46), the NIHR GOSH/UCL Biomedical Research Centre Advanced Therapies for Structural Malformations and Tissue Damage pump-prime funding call (grant n. 117DS18), the Engineering and Physical Sciences Research Council (EPSRC, grant n. EP/N02124X/1), and the European Research Council (ERC-2017-StG-757923). This report incorporates independent research from the National Institute for Health Research Biomedical Research Centre Funding Scheme. The views expressed in this publication are those of the author(s) and not necessarily those of the Wellcome Trust, the National Health Service, the National Institute for Health Research or the Department of Health.

        \paragraph{Statement of Ethical Approval}
            The collection of patient data was approved by the UK Research Ethics Committee (UK REC 15/LO/0386) and Great Ormond Street Hospital for Children R\&D office (R\&D 14DS25).

        \paragraph{Declaration of Competing Interests}
            E.O.S. and A.P. currently work with Huawei Technologies Co., Ltd. They were with Imperial College London and University College London during the experiments. The other authors declare no competing interests.

        \paragraph{Author contributions statement}
            S.F. and D.J.D. initiated and designed the study. S.F. developed the algorithms, designed the data analysis and performed the experiments. B.K. and M.J.C. validated the methodology and helped design the study. A.J.R., D.J.D., N.u.O.J. and S.S. validated the results. E.O.S., L.S.v.d.L., and A.P. pre-processed the data. S.F. wrote the paper in conjunction with A.J.R., who wrote the clinical introduction and discussion. B.K., D.S., R.K., S.S., N.u.O.J., D.J.D, and M.J.K. provided manuscript feedback.

    \bibliographystyle{unsrt}
    \bibliography{bibliography}

\end{document}


\flushbottom
    \maketitle
    
    \section*{Analysis of Complex Cases}
        
        \begin{figure*}
            \centering
            \includegraphics[width=\textwidth]{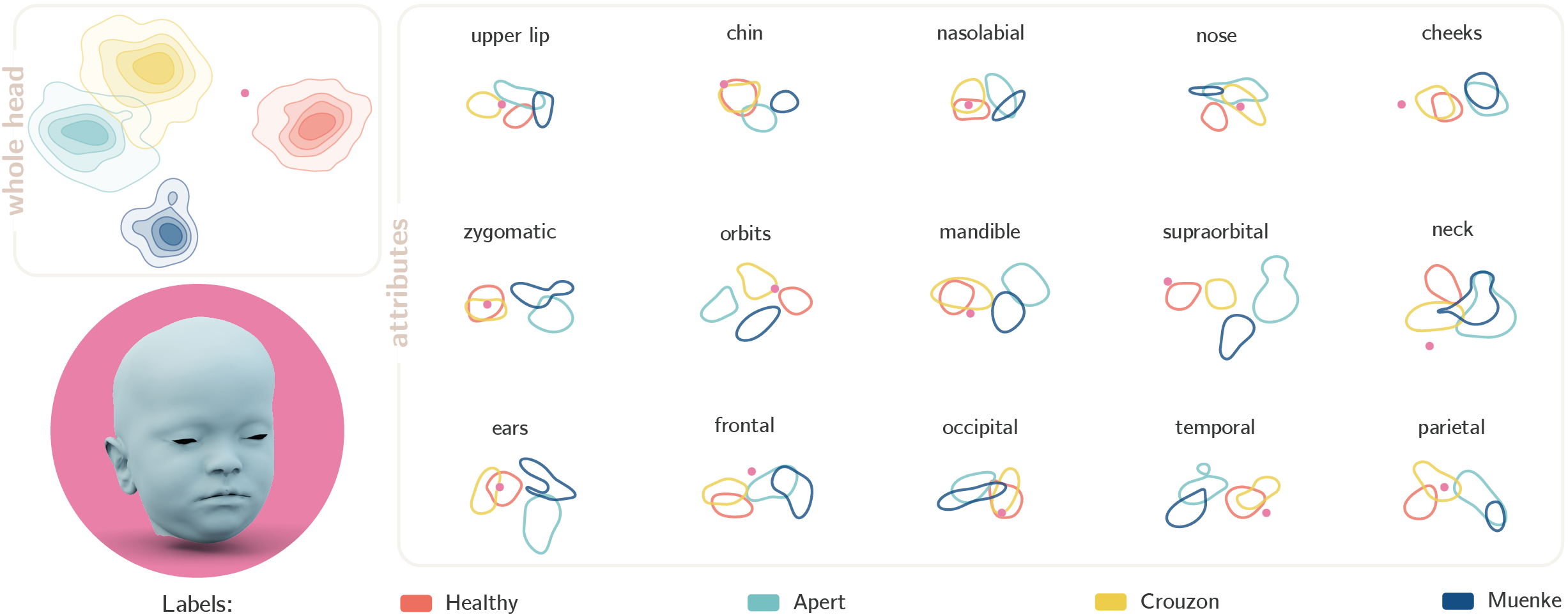} 
            \caption{\normalfont Analysis of an atypical Crouzon patient, represented in the pink circle. \textit{Top-left:} embedding of the whole latent representation of the patient's head (pink dot) in comparison to the latent distribution of each syndrome. \textit{Right:} attribute-specific embedding of the latent subsets controlling each different anatomical region. The pink dots represent the patient's attributes, while the iso-contour lines show the healthy and syndromic distributions at one standard deviation.}
            \label{fig:atypical}
        \end{figure*}    
        
        \noindent We also tested our method on an atypical Crouzon patient from O’Sullivan et al. (2021). This two-year-old patient was judged to be clinically healthy by an expert, but was shown to have Crouzon syndrome through genetic testing. Like O’Sullivan et al. (2021), our method also outperforms expert diagnosis in this instance, and correctly identifies the syndrome. In addition, disentanglement of the facial features provides greater insight into what makes this case unusual and which attributes contributed most heavily in influencing the correct classification.
        
        After encoding the patient's head with SD-VAE, QDA correctly classifies its latent representation as Crouzon. In Fig.~\ref{fig:atypical}, we report the LDA embeddings of the whole latent and of the subsets controlling each different attribute. When the whole head is considered, the latent embedding of the patient falls between the healthy and Crouzon distributions. 
        In particular, it appears to be embedded closer to the Healthy distribution, which is consistent with the expert diagnosis. Note that QDA is able to correctly classify the patient because it operates on the 75-dimensional latent vector, while the LDA embedding operation inevitably loses some of the useful information that should place this patient closer to the Crouzon distribution. 
        With our method we can also analyse the embeddings of each different anatomical sub-unit. The shape of the periorbital regions are usually particularly valuable when identifying healthy subjects and diagnosing syndromic craniosynostosis (see Results Section). Nevertheless, in this patient, the shape of the supraorbital and malar regions appear healthy, while the embeddings of the orbits fall on the edge of the Crouzon distribution and closer to the Healthy population. The features clearly contributing to the correct classification of this subject are the nose, cheeks, neck, and parietal region. The subtleties in these attributes demonstrate how challenging clinical diagnosis can be, especially when considering phenotypic variation.
        As the embeddings of the other attributes are ambiguous, it may be that, as for the entire latent, some useful information was lost during the per-attribute LDA-based embeddings. However, in this case, dimensions are reduced from $5$ to $2$ and the amount of information lost should be relatively small. We therefore consider these ambiguities to be mostly determined by the attributes' shape and the patient's atypical presentation.  

    \section*{Additional Procedure-Specific Latent Interpolations}

\begin{figure*}
            \centering
            \includegraphics[width=\textwidth]{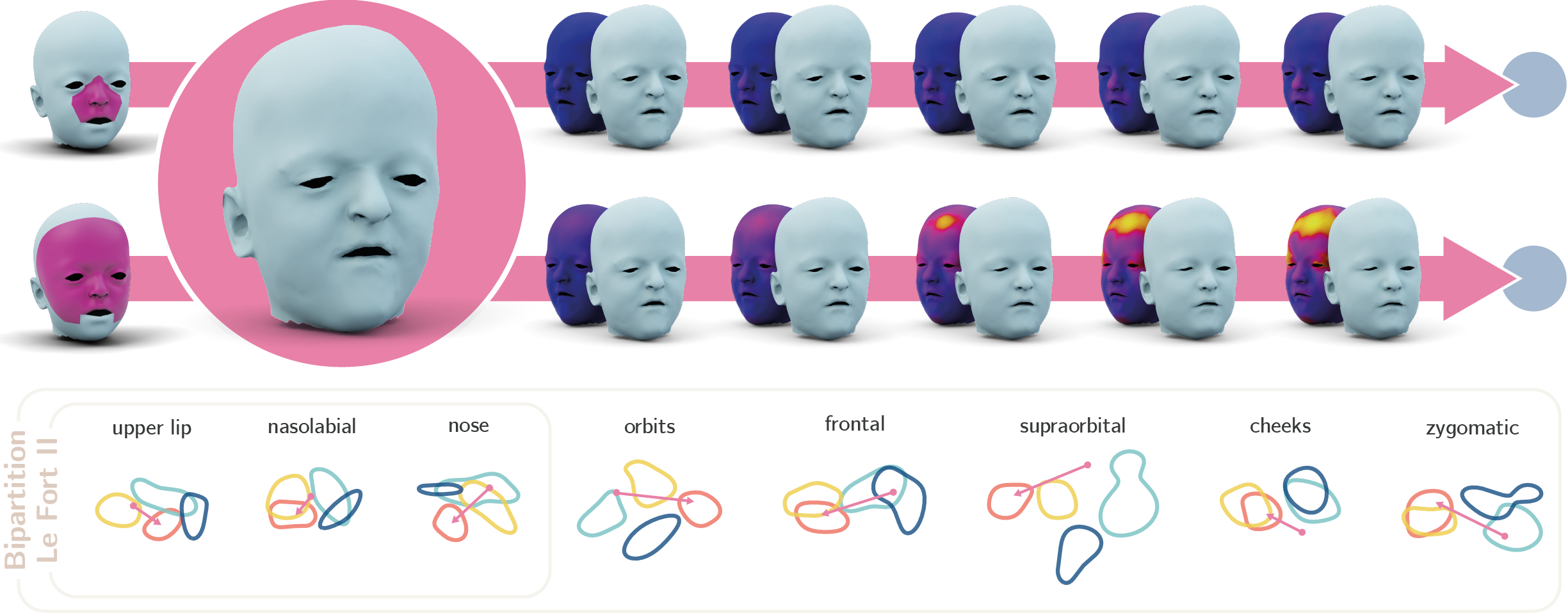} 
            \caption{\normalfont Bipartition and Le Fort II latent interpolations for an Apert patient. \textit{Top:} the patient's head is represented in the pink circle. On its left, the anatomical sub-units affected by the procedures are highlighted. On the right of the pink circle, the shapes generated during the procedure-specific interpolation are rendered alongside a displacement map showing the magnitude of the displacements (blue corresponds to no displacement and yellow to displacements of more than $10$ mm). Both interpolations stop at the blue points represented in Fig.~\ref{fig:apert_muenke_global_interp}. \textit{Bottom:} embeddings of the latent interpolations for the different subsets of latent variables controlling the anatomical sub-units affected by the two procedures.}
            \label{fig:apert_planning}
        \end{figure*}

\begin{figure*}
            \centering
            \includegraphics[width=\textwidth]{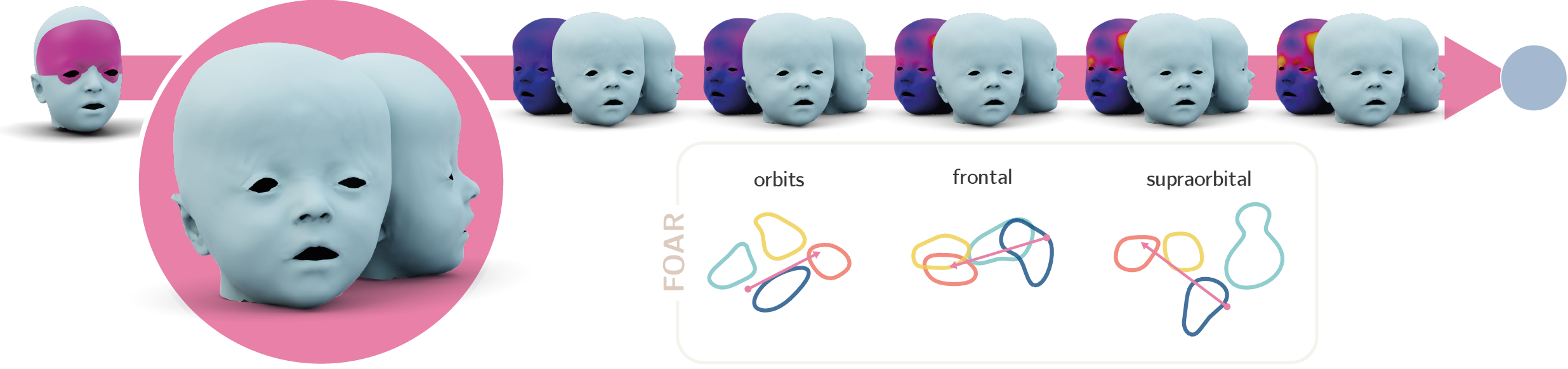} 
            \caption{\normalfont FOAR latent interpolations for a Muenke patient (see caption of Fig.~\ref{fig:apert_planning} for more details).}
            \label{fig:muenke_planning}
        \end{figure*}

\begin{figure*}
            \centering
            \includegraphics[width=\textwidth]{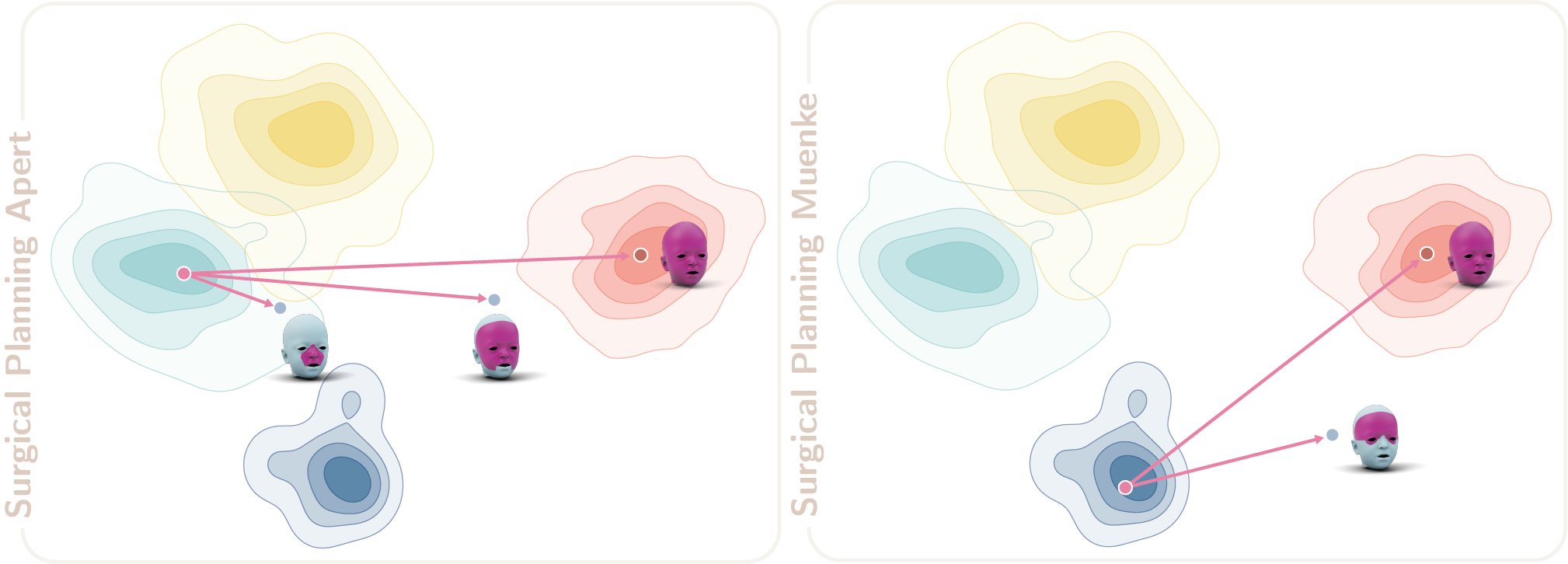} 
            \caption{\normalfont Global interpolation trajectories determined by the procedure-specific latent-interpolations for the Apert (Fig.~\ref{fig:apert_planning}) and Muenke (Fig.~\ref{fig:muenke_planning}) patients. The ideal interpolation trajectory, where the whole latent is interpolated towards $\mathbf{z}^H_\mu$, is displayed for comparison.}
            \label{fig:apert_muenke_global_interp}
        \end{figure*}

        Similarly to the \textit{Evaluating the effects of different surgical approaches} Section, we report the evaluation of surgical approaches for an Apert (Fig.~\ref{fig:apert_planning}) and a Muenke patient (Fig.~\ref{fig:muenke_planning}). As we undertake latent interpolation on the Apert patient for the regions influenced by frontofacial surgery, we can see that the progression towards the centre of the healthy population mimics mid-facial bipartition surgery, which is specifically used to address the deformity found in Apert syndrome. The aims are to correct both hypertelorism and the mid-facial biconcavity by improving the projection of the mid face and reducing the inter-orbital distance, which are displayed here. In contrast, for the Le Fort II advancement, the changes are more subtle but again address the disproportion in the middle facial third.
        The latent interpolation for the Muenke patient simulates how fronto-orbital remodelling (FOAR) could be used to address the brachycephaly commonly seen in patients with Muenke syndrome. Projection of the shape towards the mean achieves a more typical head shape, often the primary goal of surgery. As it can be observed in Fig.~\ref{fig:apert_muenke_global_interp}, all the simulated procedures bring the head shape of the patient closer to the healthy distribution.

    \section*{Spectral Analysis}
\begin{figure*}[t]
            \centering
            \includegraphics[width=\textwidth]{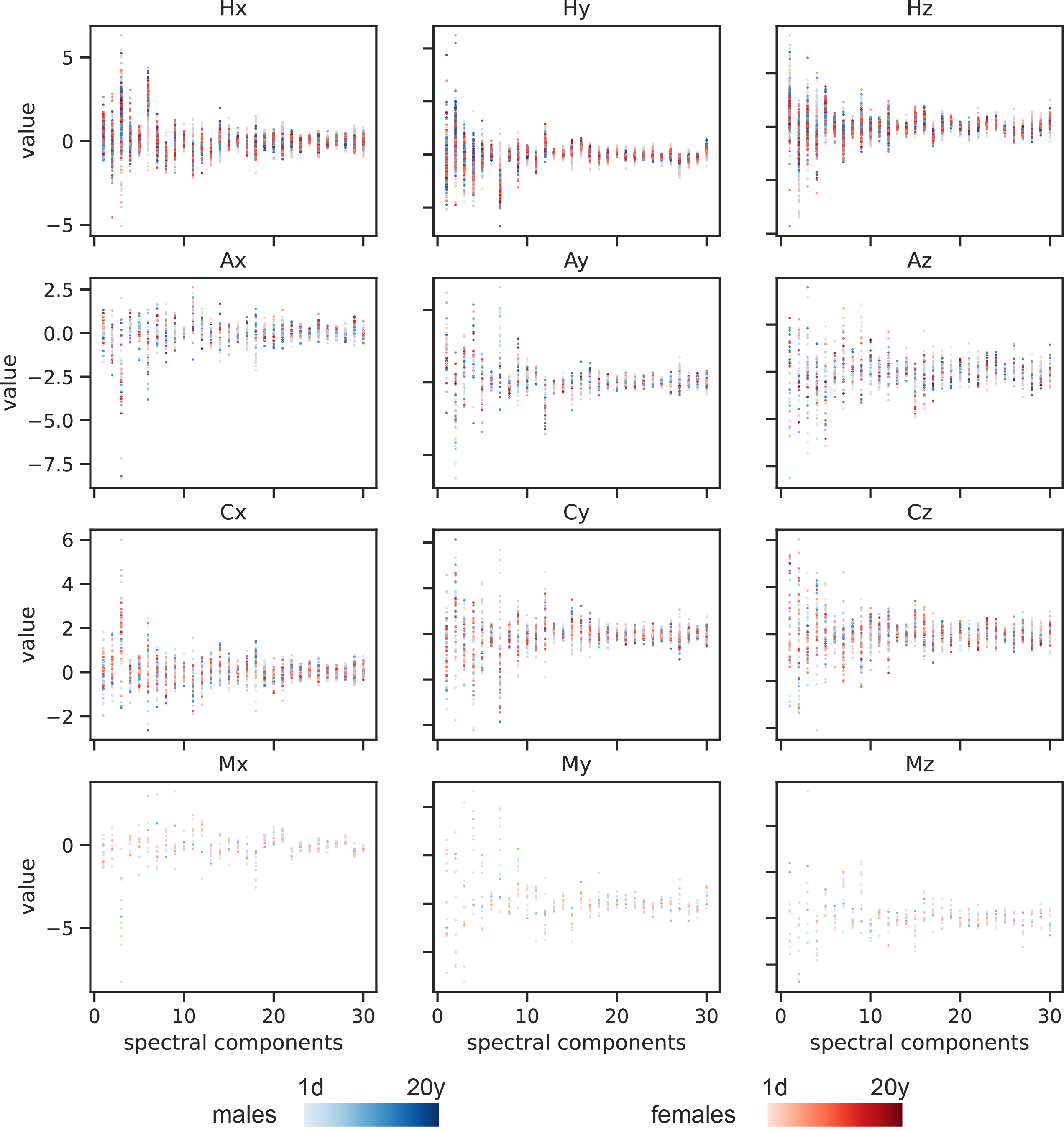} 
            \caption{\normalfont Spectra of real subjects. Each row depicts the spectra plots for the \textit{x, y}, and \textit{z} vertex coordinates of subjects within the same class. We identify these plots with Lx, Ly, and Lz, where L is the class label, with L = H for Healthy, A for Apert, C for Crouzon, and M for Muenke. A blue colourmap proportional to the age is used to plot male subjects. A red colourmap proportional to the age is used to plot female subjects. It can be observed how darker colours seem to cluster together, while red and blue markers appear scattered across most spectral components. Fig.~\ref{fig:spectral_age} and Fig.~\ref{fig:spectral_sex} better highlight this behaviour.}
            \label{fig:spectral_all}
        \end{figure*}

\begin{figure*}[t]
            \centering
            \includegraphics[width=\textwidth]{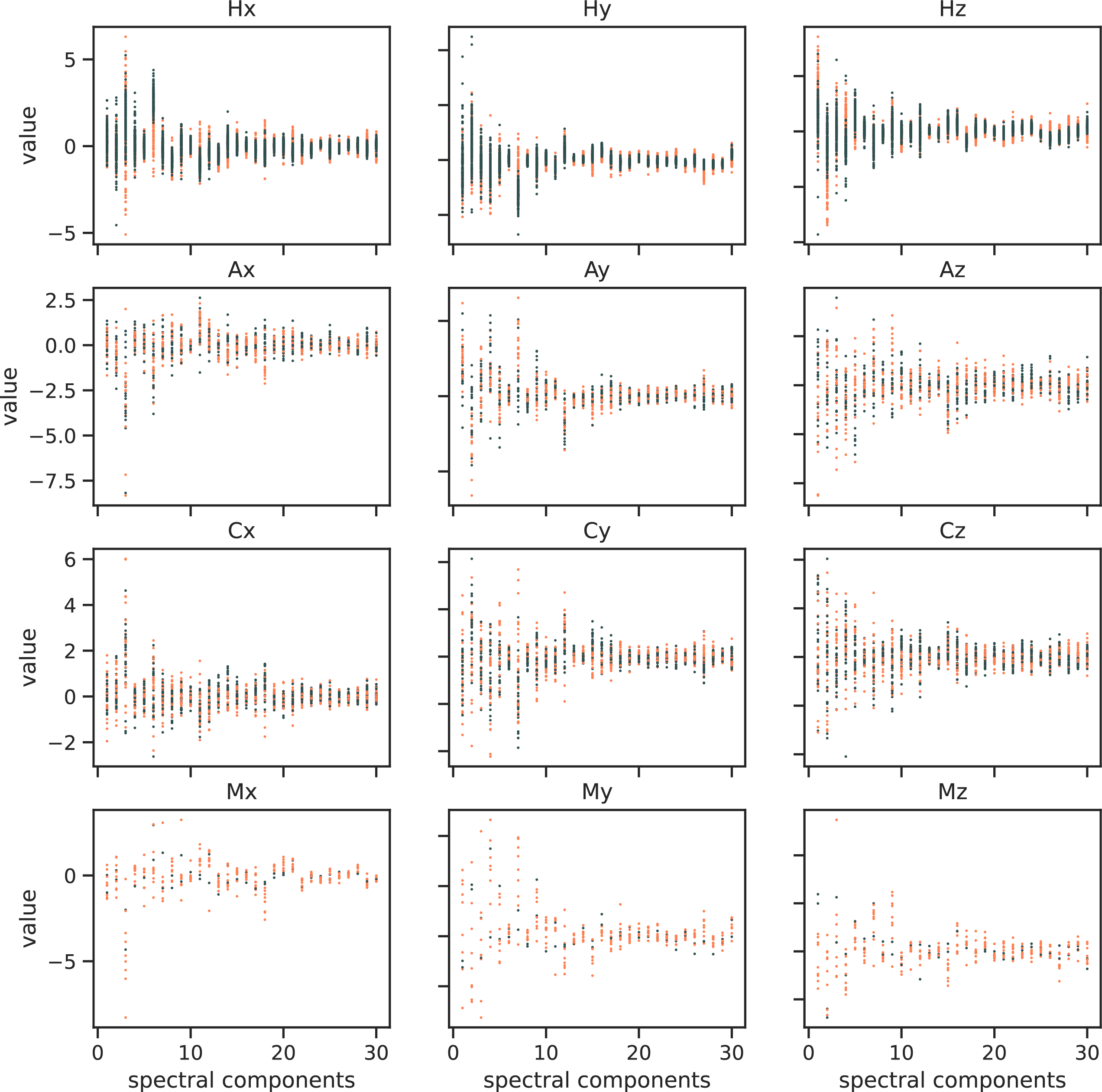} 
            \caption{\normalfont Spectra of real subjects according to their age-group. Each row depicts the spectra plots for the \textit{x, y}, and \textit{z} vertex coordinates of subjects within the same class. We identify these plots with Lx, Ly, and Lz, where L is the class label, with L = H for Healthy, A for Apert, C for Crouzon, and M for Muenke. If the age of the subject is in the $[0,4)$ years-old range, its spectra is plotted in teal. If the age of the subject is in the $[4,20]$ years-old range, its spectra is plotted in orange. Setting an age related threshold and using two distinct colours to represent the two different age ranges clearly highlights the presence of age-related clusters. This corroborates the initial findings inferred observing Fig.~\ref{fig:spectral_all} and motivates us to select only mesh pairs within the same age group during the data augmentation.}
            \label{fig:spectral_age}
        \end{figure*}

\begin{figure*}[t]
            \centering
            \includegraphics[width=\textwidth]{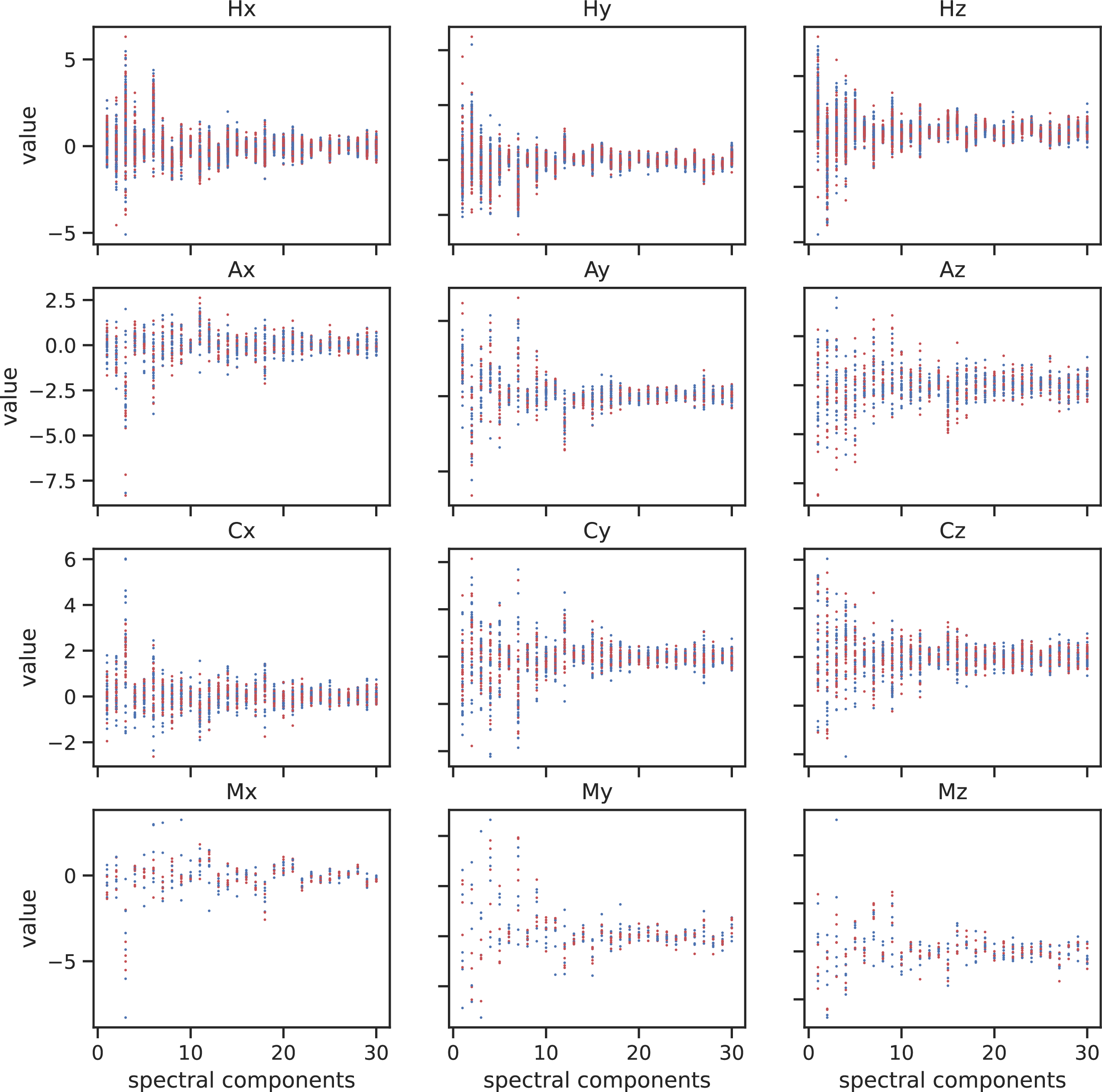} 
            \caption{\normalfont Spectra of real subjects according to their sex. Each row depicts the spectra plots for the \textit{x, y}, and \textit{z} vertex coordinates of subjects within the same class. We identify these plots with Lx, Ly, and Lz, where L is the class label, with L = H for Healthy, A for Apert, C for Crouzon, and M for Muenke. Male subjects are plotted in blue and females in red. Similarly to what previously observed in Fig.~\ref{fig:spectral_all}, the spectral components of male and female subjects are mostly mixed. Therefore, given the paucity of available data, we do not consider the subjects' sex when selecting mesh pairs for the data augmentation. The pairs can indeed represent subjects of both sexes.}
            \label{fig:spectral_sex}
        \end{figure*}

        \noindent As detailed in the Results Section, before augmenting the data with our novel spectral interpolation technique, we analyse the spectral components of all meshes in the dataset collected by O’Sullivan et al. (2021). As mentioned in the Methods Section, spectral components of each mesh are obtained by applying the Fourier transform to the vertex positions ($\mathbf{\hat{X} = U^T X}$). Although the Laplacian eigendecomposition is computed with $k=1,000$ eigenvalues and eigenvectors, our augmentation technique interpolates only the first $30$ spectral component. Therefore, we analyse only the first $30$ rows of $\mathbf{\hat{X}}$, which is a $1,000 \times 3$ matrix whose rows represent the different spectral components and whose columns are related to the $x, y, z$ coordinates of the vertices.
        The $30$ components of all meshes in the dataset are separately plotted for $x, y, z$ and for the different classes (Healthy, Apert, Crouzon, Muenke). 
        Visually inspecting the spectra provides useful insights on the differences and similarities between the different meshes in the dataset. Even if there is no intuitive interpretation of the values obtained Fourier-transforming vertex positions, identifying clusters in the spectral domain can prevent the creation of ambiguous or unrealistic head shapes. 
        In Fig.~\ref{fig:spectral_all} we plot the spectra assigning to each subject a colour proportional to the age. To differentiate between male and female subjects, colourmaps in the shades of blue and red are used respectively. To better highlight the presence of age-related clusters, in Fig.~\ref{fig:spectral_age} spectra are plotted in teal if the age of the patient is within the $[0,4)$ years-old range and in orange if the age is in the $[4,20]$ years-old range. Similarly, in Fig.~\ref{fig:spectral_sex} we plot the spectra in blue for males and in red for females. However, only a few sex-related clusters can be identified. For this reason, only the age-group is considered during the sample selection for the data augmentation. 

